\documentclass[12pt,preprint]{aastex}
\shorttitle{RELATIVISTIC PHASE SHIFT IN PULSARS}
\shortauthors{R. T. Gangadhara}
\received{}
\input epsf.tex
\usepackage{graphicx}
\def\eps{\hat \epsilon}

\begin{document}
\title{On the method of estimating emission altitude from relativistic
       phase shift in pulsars}

\author{R. T. Gangadhara }
\affil{Indian Institute of Astrophysics, Bangalore -- 560034, India\\
\email{ganga@iiap.res.in}}
\begin{abstract}
In the preprint astro-ph/0411161 (hereafter G04b), the term $-(\Omega r/c) 
\sin\theta'\cos\Theta$ is missing in the equation (13) for $\cos\psi.$ 
So, in this is supplementary preprint I give the corrected expression for 
$\cos\psi$ and the revised derivation of aberration phase shift. 
\end{abstract}
\section{PHASE SHIFT OF RADIATION EMITTED BY A PARTICLE (BUNCH)}
Consider the particle velocity given by
\begin{equation}\label{totv1}
{\bf v} = \kappa c\, {\hat b}_{\rm 0t}+{\bf v}_{\rm rot}.
\end{equation}
By substituting for ${\bf v}_{\rm rot}$ (eq. 3 in G04b) into equation~(\ref{totv1})
we obtain
\begin{equation}\label{totv}
{\bf v} = (\kappa c+\Omega r \sin\theta'\cos\Theta) \, {\hat b}_{\rm 0t}+\Omega r \sin\theta'\sin\Theta\,
           \eps_\perp \, .
\end{equation}

By assuming $\vert{\bf v}\vert\sim c,$ from equation~(\ref{totv}) we obtain the parameter
\begin{equation}\label{kap}
\kappa = \sqrt{1-\left(\frac{\Omega r}{c}\right)^2 \sin^2\theta'\sin^2\Theta}-
    \frac{\Omega r}{c} \sin\theta'\cos\Theta\, .
 \end{equation}

Using the equations~(3) and (4) given in G04b, we can solve for $\eps_\perp$ and obtain\begin{equation}\label{eps_perp}
\eps_\perp = \frac{\hat\Omega\times\hat r}{\sin\theta'\sin\Theta}-\cot\Theta\,\hat b_{\rm 0t}~.
\end{equation}
Let $\psi$ be the angle between the rotation axis and $\bf v,$ then we have
\begin{eqnarray}
\cos\psi = \hat \Omega . {\hat v} & = &
\cos\zeta \left (\sqrt{1-\left(\frac{\Omega r}{c}\right)^2 \sin^2\theta'\sin^2\Theta}-\frac{\Omega r}{c}
\sin\theta'\cos\Theta\right)~\nonumber \\
& = & \kappa\, \cos\zeta~ ,
\label{psi}
 \end{eqnarray}
where ${\hat v}={\bf v}/\vert {\bf v}\vert .$
For $r/r_{_{\rm LC}}\ll 1,$ it reduces to $\psi\sim\zeta.$

\subsection{\it Aberration Angle}

If $\eta$ is the aberration angle, then we have
\begin{eqnarray}
\cos\eta & = & {\hat b}_{\rm 0t}\cdot {\hat v} =\kappa+\frac{\Omega r \sin\theta'
               \cos\Theta}{c}~,\label{coseta}\\
\sin\eta & = & \eps_\perp \cdot {\hat v} = \frac{\Omega\, r}{c}\sin\theta'\sin\Theta~.
\label{sineta}
\end{eqnarray}
Therefore, from equations~(\ref{coseta}) and~(\ref{sineta}), we obtain
\begin{equation}\label{eta1}
\tan\eta = \frac{\Omega r}{c}{\sin\theta'\sin\Theta\over
\sqrt{1-(\Omega r/c)^2 \sin^2\theta'\sin^2\Theta}}.
\end{equation}
Hence the radiation beam, which is centered on the direction of $\bf v,$ gets tilted (aberrated) 
with respect to ${\hat b}_{\rm 0t}$ due to rotation.  

For $\Omega r/c\ll 1$, it can be approximated as
\begin{equation}\label{eta2}
\tan\eta \approx\frac{\Omega r}{c}\sin\theta'\sin\Theta~.
\end{equation}

\subsection{\it Aberration Phase Shift}

Consider Figure~5 (G04b) in which ZAD, ZBX, ZCY and DXY are the great circles centered on the
neutron star. The small circle ABC is parallel to the equatorial great circle DXY. The unit 
vector $\hat b_{\rm 0t}$ represents a field line tangent, which makes the angle $\zeta$ with 
respect to the rotation axis ZO. The velocity unit vector $\hat v$ is inclined by the angles
$\eta$ and $\psi$ with respect to $\hat b_{\rm 0t}$ and ZO, respectively. We resolve the vectors
$\hat b_{\rm 0t}$ and $\hat v$ into the components parallel and perpendicular to the rotation axis:
\begin{equation}\label{bt}
\hat b_{\rm 0t}=\sin\zeta~\hat b_{\rm 0t\perp}+\cos\zeta~\hat \Omega~,
\end{equation}
\begin{equation}\label{v}
\hat v=\sin\psi~\hat v_{\rm \perp}+\cos\psi~\hat \Omega~,
\end{equation}
where the unit vectors $\hat b_{\rm 0t\perp}$ and $\hat v_{\rm \perp}$ lie in the plane of small circle
ABC. Next, by solving for $\hat b_{\rm 0t\perp}$ and $\hat v_{\rm \perp}$, we obtain
\begin{equation}\label{bp}
\hat b_{\rm 0t\perp}={1\over\sin\zeta}(\hat b_{\rm 0t}-\cos\zeta~\hat \Omega)~,
\end{equation}
\begin{equation}\label{vp}
\hat v_{\rm \perp}={1\over\sin\psi}(\hat v-\cos\psi~\hat \Omega~).
\end{equation}
By taking scalar product with $\hat b_{\rm 0t\perp}$ on both sides of equation~(\ref{vp}), we obtain
\begin{equation}
\cos(\delta\phi'_{\rm abe})=\hat v_{\rm \perp}\cdot \hat b_{\rm 0t\perp}={1\over\sin\psi}(\hat v \cdot 
\hat b_{\rm 0t\perp}- \cos\psi~\hat \Omega \cdot \hat b_{\rm 0t\perp}~).
\end{equation}
Since  $\hat \Omega$ and $\hat b_{\rm 0t\perp}$ are orthogonal, we have
\begin{equation}
\cos(\delta\phi'_{\rm abe})={1\over\sin\psi}(\hat v \cdot \hat b_{\rm 0t\perp}).
\end{equation}
Using $\hat b_{\rm 0t\perp}$ from equation~(\ref{bp}) we obtain
\begin{equation}
\cos(\delta\phi'_{\rm abe})={1\over\sin\psi}{(\hat v \cdot \hat b_{\rm 0t}-
                            \cos\zeta~\hat v \cdot \hat \Omega)\over\sin\zeta}.
\end{equation}
By substituting for 
$\hat v \cdot \hat b_{\rm 0t}$ and $\hat v \cdot \hat \Omega$ 
from equations~(\ref{coseta}) and (\ref{psi}), we obtain
\begin{equation}
\cos(\delta\phi'_{\rm abe})={1\over\sin\psi}{(\cos\eta -
                            \cos\zeta~\cos\psi)\over\sin\zeta}.
\end{equation}
Substituting for $\eta$ again from  equation~(\ref{coseta}),
we obtain
\begin{equation}
\cos(\delta\phi'_{\rm abe})={1\over\sin\zeta\sin\psi}\left(\kappa+{\Omega r\over c} \sin\theta'\cos\Theta - \cos\zeta\cos\psi \right ).
\end{equation}
By substituting for $\kappa$ from equation~(\ref{psi}), we obtain
\begin{equation}
\cos(\delta\phi'_{\rm abe})={1\over\sin\zeta\sin\psi}\left(\frac{\cos\psi}{\cos\zeta}+
\frac{\Omega r}{c} \sin\theta'\cos\Theta - \cos\zeta\cos\psi \right ).
\end{equation}
It can be further reduced to
\begin{equation} \label{dphiabec1} 
\cos(\delta\phi'_{\rm abe})  = \tan\zeta\cot\psi+\frac{\Omega r}{c}\frac{\sin\theta'\cos\Theta}{\sin\zeta\sin\psi}
\end{equation}
 
Next, by taking a scalar product with $\eps_\perp$ on both sides of equation~(\ref{vp}), 
we obtain
\begin{equation}\label{sindph}
\sin(\delta\phi'_{\rm abe})=\hat v_{\rm \perp}\cdot \eps_\perp =
{1\over\sin\psi}(\hat v \cdot \eps_\perp-\cos\psi~\hat \Omega \cdot \eps_\perp~).
\end{equation}
Since $\hat \Omega \cdot (\hat\Omega\times\hat r)=0,$ from equation~(\ref{eps_perp}) we obtain
\begin{equation}
\hat\Omega\cdot\eps_\perp=-\cot\Theta\, (\hat\Omega\cdot{\hat b}_{0t}).
\end{equation}
We know from equation~(\ref{bt}), $(\hat\Omega\cdot{\hat b}_{0t})=\cos\zeta.$ Therefore, we have
\begin{equation}\label{omeps}
\hat\Omega\cdot\eps_\perp=-\cot\Theta\, \cos\zeta.
\end{equation}
By substituting for $\hat v \cdot \eps_\perp$ from equation~(\ref{sineta}), and for
$\hat\Omega\cdot\eps_\perp$ from equation~(\ref{omeps}) into equation~(\ref{sindph})
we obtain
\begin{eqnarray}
\sin(\delta\phi'_{\rm abe}) &= & {1\over\sin\psi}(\sin\eta+\cos\zeta\cos\psi\cot\Theta)\nonumber \\
& = & {\Omega r\over c}{\sin\theta'\sin\Theta\over\sin\psi} +\cos\zeta\cot\psi\cot\Theta~.
\label{dphiabes1}
\end{eqnarray}
Thus, we obtain from equations~(\ref{dphiabec1}) and (\ref{dphiabes1}):
\begin{equation}
\tan(\delta\phi'_{\rm abe}) = 
  {(\Omega r/c)\sin\theta'\sin\Theta  +\cos\zeta\cos\psi\cot\Theta\over 
\tan\zeta\cos\psi+(\Omega r/c)(\sin\theta'\cos\Theta/\sin\zeta) }~.
\end{equation}
\end{document}